\documentstyle[preprint,aps,prb,epsf]{revtex}

\newcommand{\mb}[1]{ { \mbox{\boldmath{$#1$}}}  }

\begin{document}
\draft

\title{Superconductivity In Disordered Sr$_2$RuO$_4$}

\author{G. Litak}

\address{Department of Mechanics, Technical University of Lublin, 
20-618 Lublin, Poland}

\author{J.F. Annett, B.L. Gy\"{o}rffy}
\address{H.H. Wills Physics Laboratory, University of Bristol, Tyndall
Ave,
 Bristol,
BS8 1TL, UK}

\date{\today}
\maketitle

\begin{abstract}
We discuss the influence of disorder on the critical temperature $T_C$ of
a $p$-wave
superconductor. To describe disordered Sr$_2$RuO$_4$ we use extended Hubbard model
with random site energies treated in the Coherent Potential Approximation.
\end{abstract}

\section{Introduction}
Recent experimental evidence suggests that the Cooper pairs in
superconducting Sr$_2$RuO$_4$ are triplets with $p$-wave
internal symmetry, as in the case of superfluid
$^3$He \cite{Mac98,Lu96,Agt97,Lit00}. 
One of characteristic features of this exotic state is the strong
influence of impurities on its superconducting properties. 
Studies of the electronic structure in 
Sr$_2$RuO$_4$ \cite{Lu96} have identified an extended van Hove
singularity 
close to the Fermi
energy $E_F$. In this note we investigate the interplay between
the van Hove singularity and disorder in a model $p$-wave superconductor.
\section{The Model}
We consider a simple extended Hubbard Hamiltonian \cite{Lit00,Mar99}:
\begin{equation}
\label{eq1}
H =
 \sum_{i \sigma} (\varepsilon_i-\mu) \hat n_{i\sigma} +
 \sum_{ij \sigma} t_{ij} c^+_{i \sigma}c_{j \sigma}  
+ \frac{1}{2} \sum_{ij \sigma
\sigma'} U_{ij}\hat n_{i \sigma} \hat n_{i
\sigma'} 
\end{equation}
where as usual 
$c^+_{i\sigma}$ and $c_{i \sigma}$ 
are the Fermion creation and annihilation operators for an electron on
site $i$ with spin $\sigma$,  
$\hat n_{i \sigma}$ is the number operator and 
$\mu$ is the chemical potential. Disorder is introduced into
the problem by allowing the local site energy 
$\varepsilon_i= \pm \delta/2$ to vary
randomly from site to site with equal probability. $U_{ij}$ is the attractive interaction
($ i \neq j$) between nearest sites  and $t_{ij}$ is the
hopping integral from site $j$ to site $i$ which takes nonzero values 
between nearest and
next nearest sites. In $k-$space: 
$\varepsilon_{{\small \mb k} } = \sum_{j} t_{ij} \exp{ (\imath \mb R_{ij}
\mb k)}
 = -2t (\cos k_x+ \cos k_y) - 4t' \cos k_x \cos k_y,$   
where the hopping parameter
$t'=0.45t$ as well as the band filling $n_{\gamma}/2= 0.66$
were fitted to the experimental cyclotron
masses and corresponding carriers occupations for the $\gamma$ band 
of Sr$_2$RuO$_4$, the interaction $U_{ij}/t=0.446$ was chosen to get
$T_C=1.5$K 
\cite{Mac98,Agt97,Lit00}. 
\section{Density of States and $T_C$}
 The linearized gap equation for the critical temperature $T_C$ of 
p-wave pairing 
from the Hamiltionian (\ref{eq1}) reads \cite{Lit00}:
\begin{equation}
\label{eq2}
1= \frac{|U|}{\pi} \int^{\infty}_{- \infty} {\rm d} E {\rm Tanh} 
\frac{E}{2k_BT_C }
{\rm Im} \frac{ \overline G^p_{11} (E)}{ 2 E - {\rm Tr} \mb \Sigma (E)},
\end{equation} 
where  $\overline G^p_{11} (E)$ is an averaged electron Green function
 which defines the
weighted density of states (DOS) of
$p$-wave electron states $\overline N_p( E)$:
\begin{equation}
\label{eq3}
\overline N_p(E)= - \frac{1}{\pi} \overline G^p_{11} (E) = - \frac{1}{\pi
N}
\sum_{ \small \mb k}
{\rm Im} \frac{2 \sin^2    
k_x}{E - \Sigma_{11}(E) - \varepsilon_{\small \mb k} + \mu},
\end{equation}
where
$\Sigma_{11}(E)$ is a
Coherent Potential which describe the electron
self energy in the  disordered system.

\begin{figure}[htb]
\leavevmode

\epsfxsize=5.0cm
\hspace{2.5cm}
\epsffile{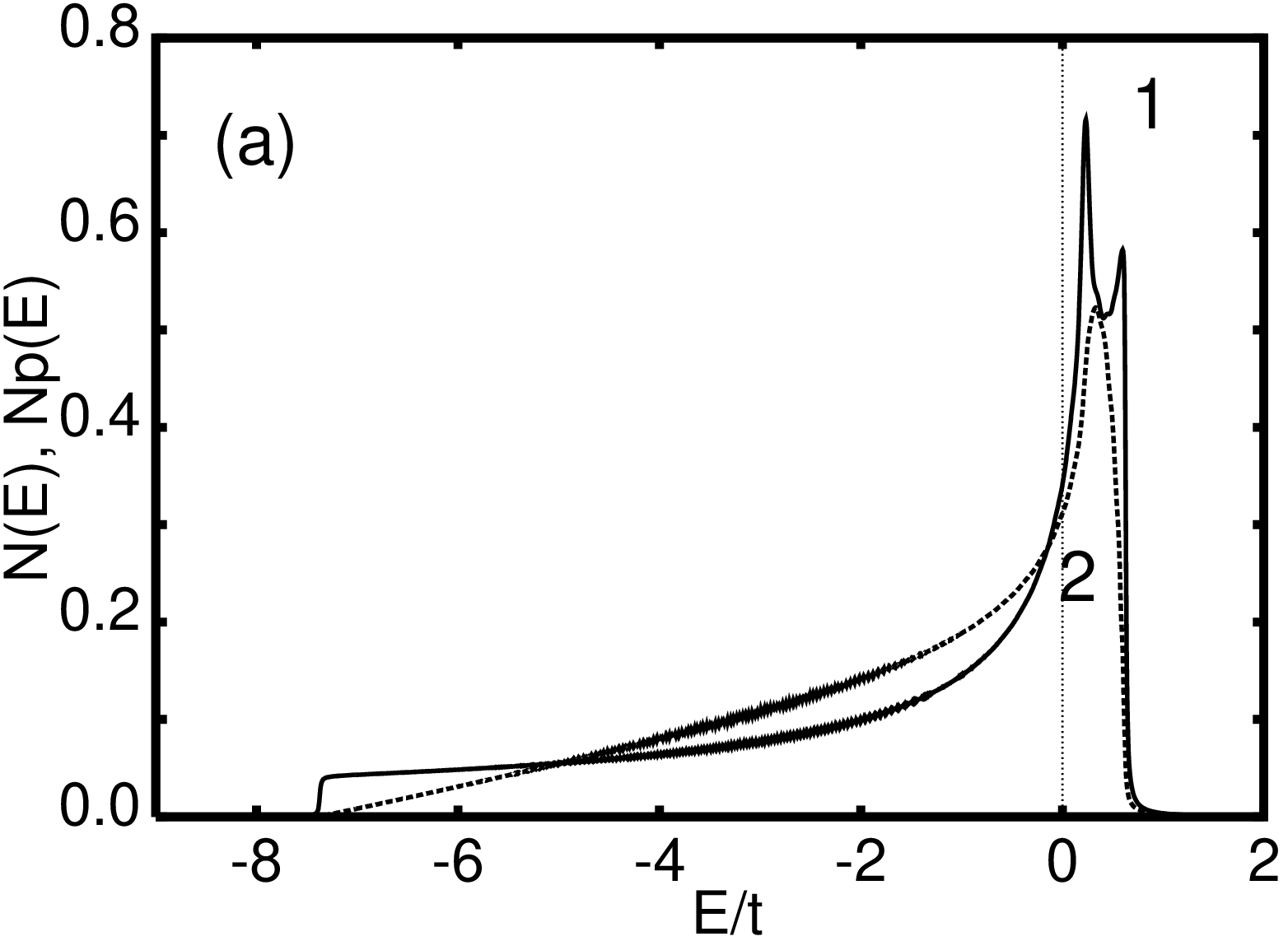}

\vspace*{-1.5cm}

\epsfxsize=5.0cm
\hspace{2.5cm}
\epsffile{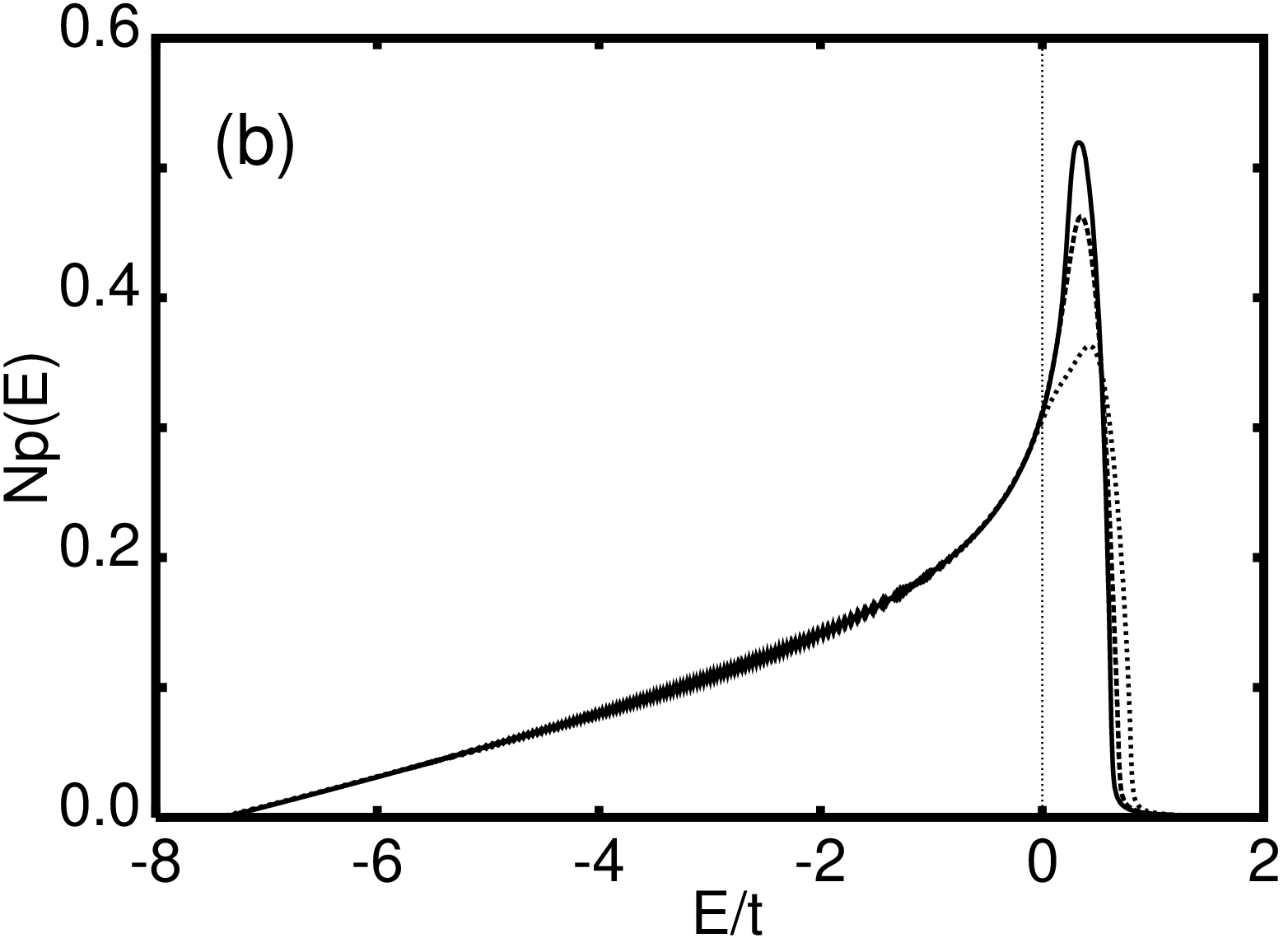}
\caption{ (a) $N(E)$ (1) and
$N_p(E)$ (2) in a clean
system, (b) $\overline N_p(E)$ for disordered system with
$\delta/t=0.0,0.3,0.6$ starting from the top line ($\mu=0$). }
\label{fig1}
\end{figure}

\begin{figure}[htb]
\leavevmode 
\epsfxsize=5.0cm
\hspace{2.5cm}
\epsffile{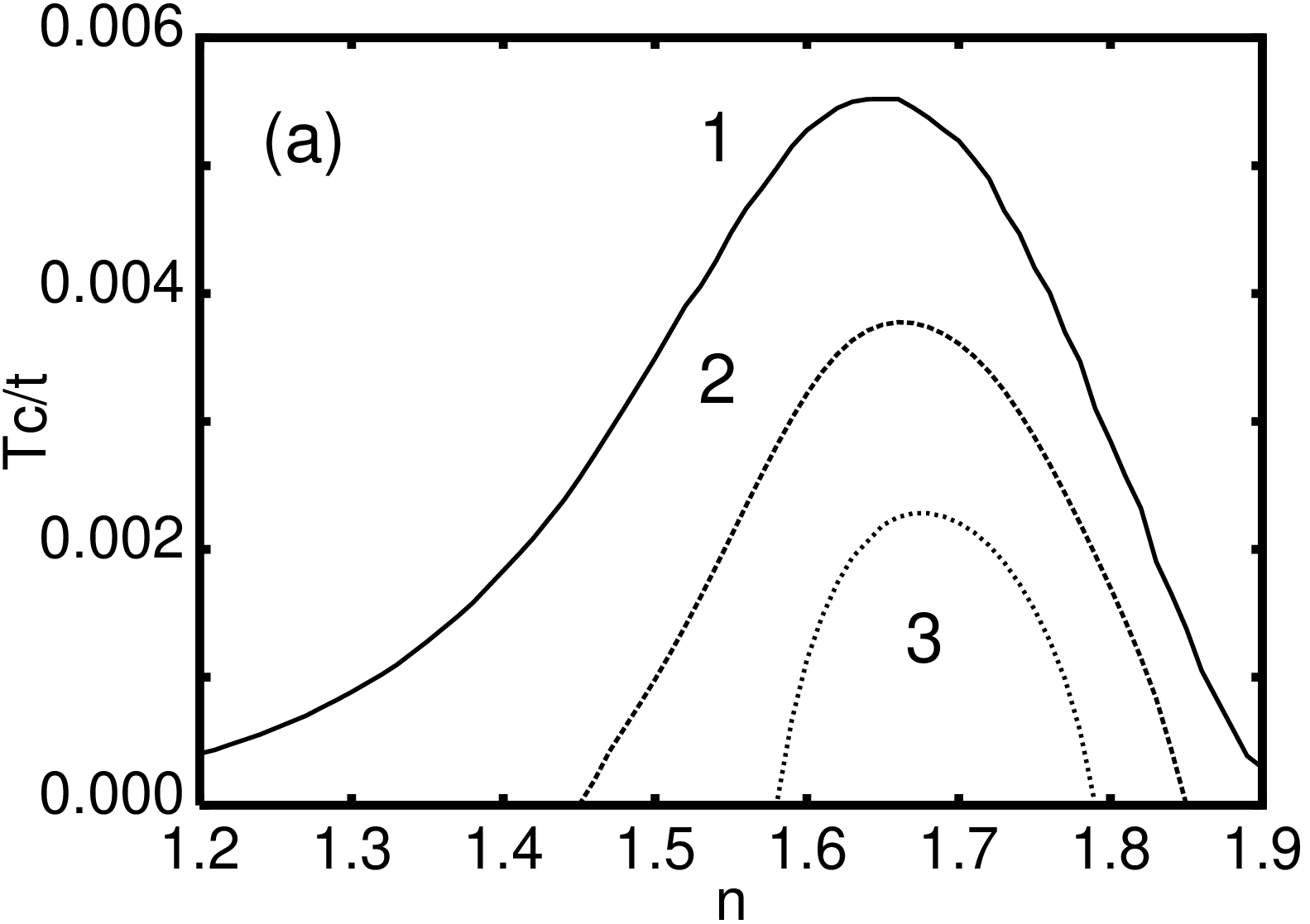}

\vspace*{-1.5cm}

\epsfxsize=5.0cm
\hspace{2.5cm}
\epsffile{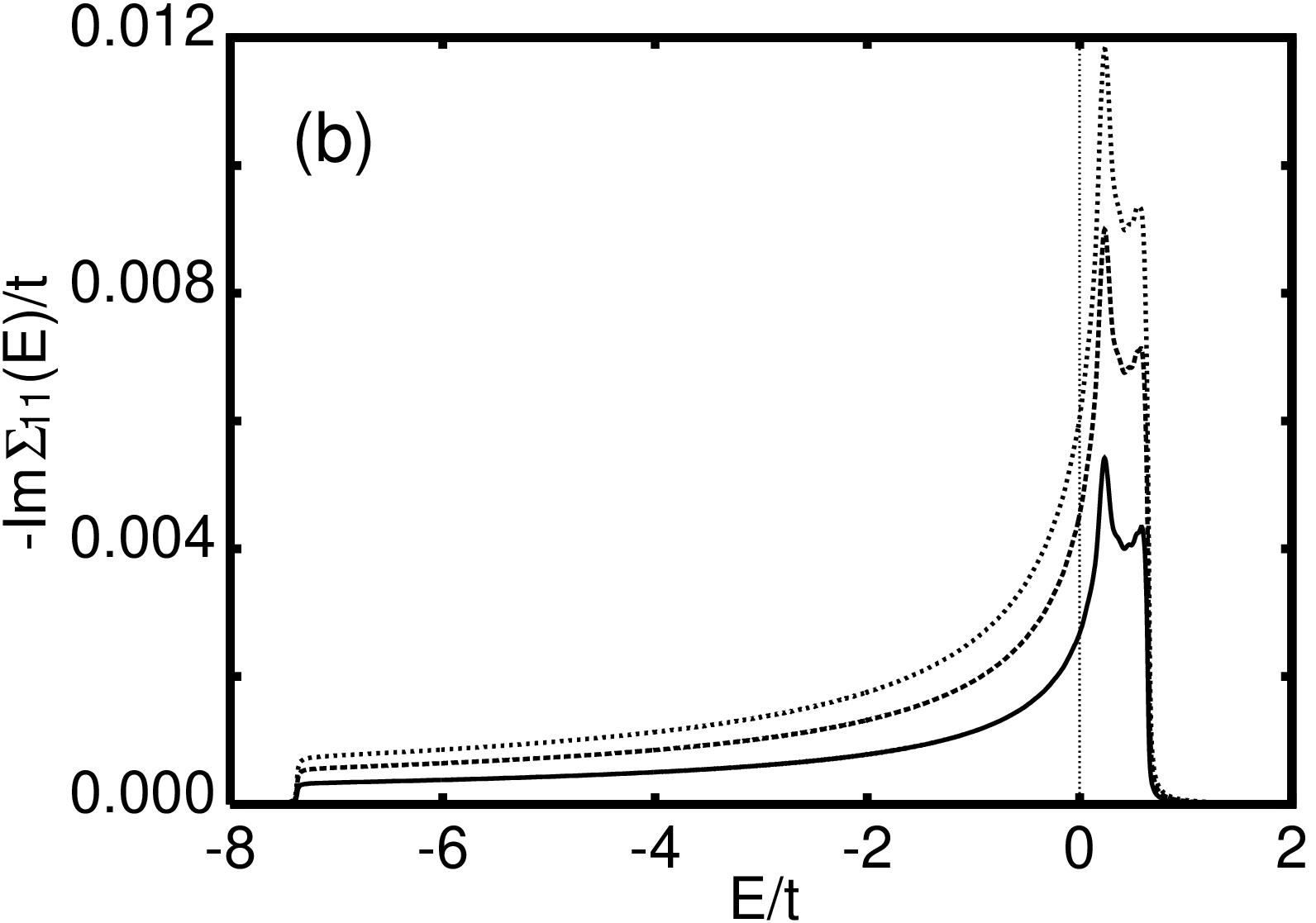}
\caption{ (a) $T_C$ as a function of $n$ for various values of
disorder potential ($\delta/t=0.0,0.10,0.13$ for curves 1, 2, 3 respectively,  (b)
Imaginary part of self energy for various values of
disordered potential $\delta/t=0.10,0.13,0.15$ starting form the bottom line
($\mu=0$).}
\label{fig2}
\end{figure}

In case of a clean system (${\rm Im} \Sigma_{11} (E) =0$)
we get a conventional gap equation with the DOS $N(E)$ substituted by
$N_p(E) $ under the integral (Eq. \ref{eq2}).
In Fig 1a we show $N(E)$ and $N_p(E)$ for the clean system. Note 
that the van Hove singularity in $N(E)$
produces the maximum in $N_p(E)$.  
The singularity in $ N_p(E)$ is smeared  by the presence of the term
${\rm sin}^2 k_x$ in
 Eq. \ref{eq3}. 
This leads to a maximum in $T_C$ with changing $n$ (Fig. \ref{fig1}a).
Equations \ref{eq2} and \ref{eq3} are influenced by  disorder by
different
effects. 
Firstly, the peak in $\overline N_p (E)$ is smeared (Fig. \ref{eq1}b)
leading 
to small
decrease of $T_C$.
The second and more interesting effect arises from Eq. \ref{eq2}, 
where $\Sigma_{11}(E)$ acts as a  pair breaker.
Using the arguments of Ref. \cite{Mar99}, Eq. \ref{eq2} can be evaluated
to yield:
\begin{equation}
\label{eq4}
{\rm ln}\left( \frac{T_C}{T_{C0}}\right) = \psi \left(\frac{1}{2}\right) -
\psi \left( \frac{1}{2} + \frac{|{\rm Im}\Sigma_{11}(0)|}{2 \pi
T_C}\right),
\end{equation}
where $T_{C0}$ denotes  the critical temperature in a clean system.
The full influence of disorder on $\overline N_p (E)$ is illustrated
in Fig. \ref{fig1}b.  Note that the position of  the maximum value    
is not affected by small disorder. On the other hand the critical 
temperature $T_C$, plotted in Fig. \ref{fig2}a is degradated strongly with 
disorder.  This is due to the pair-breaking term
$-{\rm Im} \Sigma_{11}(E)$,  which is shown in Fig. \ref{fig2}b.
\section{Summary}
When the Fermi energy is close to a Van Hove singularity
relatively weak disorder can cause very rapid
$T_C$ degradation in a $p$-wave superconductor.
The case of Sr$_2$RuO$_4$ may be an example of this phenomenon.
\begin{acknowledgments}
This work was partially founded by the Committee of Scientific Research (Poland) 
through the grant KBN 2P03B09018, and the Royal Society (UK). 
\end{acknowledgments}

\end{document}